\documentclass[aps,prb,twocolumn,amsmath,amssymb,superscriptaddress,floatfix]{revtex4}
\usepackage{dcolumn}
\usepackage{bm}
\usepackage{epsfig}
\usepackage{epstopdf}
\usepackage{graphicx}
\usepackage{grffile}
\usepackage[english]{babel}
\usepackage{bm}
\usepackage{sidecap}
\usepackage{mathptmx}
\usepackage{eqnarray,amsmath}
\usepackage{txfonts}
\usepackage[colorlinks,citecolor=blue,linkcolor=purple]{hyperref} 
\usepackage[usenames,dvipsnames,svgnames]{xcolor}
\usepackage{array}
\usepackage{romannum}
\usepackage{float}

\usepackage{lipsum}
\usepackage{natbib}

\newcommand{\tobedeleted}[1]{\textcolor{green}{#1}}
\renewcommand{\tobedeleted}[1]{\relax}

\renewcommand{\section}[1]{{\par\it #1 ---}\ignorespaces}

\begin{document}
\renewcommand{\thefigure}{\arabic{figure}}
\def\be{\begin{equation}}
\def\ee{\end{equation}}
\def\ber{\begin{eqnarray}}
\def\eer{\end{eqnarray}}

\def\kv{{\bf k}}
\def\bfr{{\bf r}}
\def\qv{{\bf q}}
\def\pv{{\bf p}}
\def\sigmav{\boldsymbol{\sigma}}
\def\tauv{\boldsymbol{\tau}}
\newcommand{\h}[1]{{\hat {#1}}}
\newcommand{\hdg}[1]{{\hat {#1}^\dagger}}
\newcommand{\bra}[1]{\left\langle{#1}\right|}
\newcommand{\ket}[1]{\left|{#1}\right\rangle}

\title{Non-Linear Spin Susceptibility in Topological Insulators}
\date{\today}

\author {Mahroo Shiranzaei}
\affiliation{School of Physics, Damghan University, P.O. Box 36716-41167, Damghan, Iran}
\affiliation{School of Physics, Institute for Research in Fundamental Sciences (IPM), Tehran 19395-5531, Iran}
\affiliation{Department of Physics and Astronomy, Uppsala University, Box 516, SE-751 21, Uppsala, Sweden}

\author{Jonas Fransson}
\affiliation{Department of Physics and Astronomy, Uppsala University, Box 516, SE-751 21, Uppsala, Sweden}

\author{Hosein Cheraghchi}
\affiliation{School of Physics, Damghan University, P.O. Box 36716-41167, Damghan, Iran}

\author{Fariborz Parhizgar}\email{fariborz.parhizgar@physics.uu.se}
\affiliation{School of Physics, Institute for Research in Fundamental Sciences (IPM), Tehran 19395-5531, Iran}
\affiliation{Department of Physics and Astronomy, Uppsala University, Box 516, SE-751 21, Uppsala, Sweden}
\affiliation{NORDITA, KTH Royal Institute of Technology and Stockholm University, Roslagstullsbacken 23, SE-106 91 Stockholm, Sweden}

\begin{abstract}
We revise the theory of the indirect exchange interaction between magnetic impurities beyond the linear response theory to establish the effect of impurity resonances in the surface states of a three dimensional topological insulator. The interaction is composed of an isotropic Heisenberg, anisotropic Ising and Dzyaloshinskii-Moriya-type of couplings. We find that all three contributions are finite at the Dirac point, which is in stark contrast to the linear response theory which predicts a vanishing Dzyaloshinskii-Moriya-type contribution. We show that the spin-independent component of the impurity scattering can generate large values of the Dzyaloshinskii-Moriya-type coupling in comparison with the Heisenberg and Ising types of coupling, while these latter contributions drastically reduce in magnitude and undergo sign changes. As a result, both collinear and non-collinear configurations are allowed magnetic configurations of the impurities.
\end{abstract}
\maketitle

\section{Introduction}
Three-dimensional topological insulators (3D TI) \cite{RevModPhys.82.3045}, materials with insulating bulk states and two-dimensional gapless surface states have attracted a huge attention during the last decade, both for their fundamentally interesting properties as well as potential applications. Fascinating novel effects such as quantum anomalous Hall effect (QAHE) \cite{Chang167} and topological superconductivity \cite{Wang52} have been observed in these materials and other unprecedented effects have been proposed in the fields of electronics and spintronics \cite{prop1, prop2}. More specifically, since the surface states of these materials follow a pure Rashba-type Hamiltonian, modifications of their band dispersion can be invoked by proximity of a ferromagnet material \cite{prop2} or magnetic impurities \cite{jonas-filling, prop-mag, prop-mag2}. In the latter case the QAHE has been experimentally observed which makes the field of dilute magnetic TIs an important area of research. 
 It worth to mention that, although this experiment has been observed in magnetic TIs by several groups, the nature of the coupling between the impurities and their alignment is still under vigorous debate.

In dilute magnetic semiconductors, the magnetic impurities mostly interact indirectly via the itinerant electrons of the host system, the so-called Ruderman-Kittel-Kasuya-Yosida (RKKY) interaction \cite{ruderman,kasuya,yosida}. This interaction allows control of the magnetic properties by tuning the electronic properties of the system, which is most desirable in the field of spintronics \cite{matsukura}. As a general role, the RKKY interaction, which is proportional to the spin susceptibility of the host material, scales with the distance $R$ between spins as $R^{-d}\sin(2k_FR)$, where $d$ is the spatial dimension and $k_F$ is the Fermi wave-vector. This long range interaction can lead to ferromagnetic (FM) or anti-ferromagnetic (AFM) ordering of the impurities. In materials with spin-orbit coupling \cite{prb.69.121303, prl.106.136802, prl.106.097201, efimkin, fariborz-mos2}, an effective Dzyalosinski-Moriya (DM) type of interaction \cite{DZYALOSHINSKY1958241, PhysRev.120.91,prl.119.027202, prl.44.1538, dmitrienko2014measuring} appears between the impurities. While the isotropic Heisenberg (H)-type interaction together with the anisotropic Ising (I)-type contribution in magnetic materials favour collinear alignment of the spins, the DM-type interaction is associated with the Hamiltonian contribution ${\bf D}\cdot ({\bf S}_i\times{\bf S}_j)$, where the vector $\textbf{D}$ defines the form of relative rotation of the spins, favors a perpendicular orientation of the spins with respect to each other. The competition between these collinear and non-collinear interactions may result in exotic phases such as skyrmions, helices, and chiral domain walls \cite{nature198, alireza, Natnano723}. 
The RKKY interaction in TI has been studied widely \cite{prl.106.136802, prl.106.097201, PhysRevB.89.235316, mhf} and both collinear and non-collinear terms were reported. The importance of this interaction in TI is that the magnetic moments can couple to each other upto many nanometers in contrast to many Angstroms in magnetic semiconductors\cite{wray2011topological}. Since these terms can be tuned by changing the electronic doping and the distance between the impurities, it was proposed to deposit magnetic impurities on TIs in any desirable lattice structure \cite{prl.106.097201} or random distribution \cite{prl.106.136802} and study the resulting spin model. 
However, the fact that the DM-type interaction vanishes at the Dirac point makes realization of exotic phases such as skyrmions challenging \cite{nagaosa302}.

In Dirac materials, such as 3D TI, it has been shown that magnetic and non-magnetic impurities generate local resonances near the Dirac point \cite{biswas}. The existence of these resonances becomes more prominent when they emerge at forbidden energies near the band gap \cite{mfjh} or at low density of electron states (DOS) near the Dirac point \cite{biswas}. Note that a magnetic impurity comprises both a magnetic and a non-magnetic scattering potential. While the former potential generates both electron and hole resonance peaks located symmetrically around the Dirac point, the latter breaks the electron-hole symmetry and creates only an electron or hole resonance, depending on whether it is attractive or repulsive.  
Recent studies suggest that the gap induced by magnetic impurities may be destroyed by the accompanied non-magnetic scattering \cite{jonas-filling}. Besides, notwithstanding the peak according to the potential scattering is a universal feature of Dirac materials \cite{sasha-dirac,PhysRevB.94.075401}, the effect of magnetic term would differ in different materials with respect to their spin properties. 
The effect of impurity resonances on indirect spin-spin coupling in two-dimensional (2D) materials has been investigated recently \cite{samir, datta, mishchenko14}, however, restricted to spin-degenerate materials. Although the distances between magnetic impurities in dilute magnetic TIs are sufficiently large to suppress direct interaction between them, it is not larger than that their induced impurity states have an influence on their indirect interaction. This observation and the importance of the impurity states near the vanishing DOS at the Dirac point motivate the present Rapid Communication.

Here, we investigate the effect of the impurity resonances in the TI surface states on the RKKY interaction. First, we extend the formalism introduced in Ref. \citenum{datta} and calculate the spin susceptibility beyond the linear response theory and subsequently investigate the effects of both non-magnetic and magnetic scattering potentials on the RKKY interaction. We show that the non-magnetic scattering potential enhances the electron density near the Dirac point, which significantly modifies the properties of the interaction. In particular, we find a quadratic spatial decay in contrast to the cubic obtained in linear response. Moreover, the DM-type interaction becomes finite and non-negligible while the both H and I- types of coupling are reduced and even their sign change in some range of parameters. We show, furthermore, that our findings are not restricted to the Dirac point but is important at finite doping. Finally, we present the application of our results to the final phase of two impurities on the surface of TI.

\section{Theoretical modeling} 
The surface states of the $ 3 $D TI around the $\Gamma$ point can be described by the effective Hamiltonian \cite{prb82, NatPhys584, hamiltonian, effective} $H_0=\hbar v_F [\textbf{k}\times\hat{\bf z}]\cdot\boldsymbol{\sigma}$, where $\boldsymbol{\sigma}$ denotes the vector of the Pauli matrices corresponding to the real spin, $\textbf{k}$ is the momentum, and $v_F$ presents the Fermi velocity. We, furthermore, model the impurity at ${\bf r}_0$ by $H_\text{imp}=U\delta({\bf r}-{\bf r}_0)$, where $U=u\sigma_0+\textbf{m}\cdot\boldsymbol{\sigma}$ contains both the non-magnetic ($u$) and magnetic (${\bf m}$) scattering potentials. The latter, relates to the spin of the impurity, ${\bf S}$, via ${\bf m}=\hbar J_c {\bf S}/2$, where $J_c$ is the coupling constant between impurity and itinerant electron spins.


We approach the RKKY interaction beyond linear response theory, by identifying the local magnetization ${\bf M}({\bf r})= \boldsymbol{\chi}({\bf r},{\bf r}^\prime) \cdot {\bf m}({\bf r}')$ where $\boldsymbol{\chi}$ is the susceptibility tensor and ${\bf m}$ indicates the magnetic scattering potential given in $H_\text{imp}$.
By using this relation together with the definition of magnetization based on the spin local density of states (LDOS), ${\bf M}(\textbf{r},\varepsilon)=-{\Im\, \text{Tr}\;}\big[\boldsymbol{\sigma}{\bf G}({\rm r},{\rm r};\varepsilon)\big]/2\pi$, we capture the effect of impurity states in the spin susceptibility tensor. 
Here, ${\bf G}({\rm r},{\rm r};\varepsilon)$ is the on-site perturbed Green's function (GF) which can be obtained by using the $T$-matrix approach \cite{biswas, jonas-filling} as below
\begin{align}
\label{eq:GreenT}
{\bf G}({\bf r},{\bf r}';\varepsilon)=&
	{\bf G}_0({\bf r}, {\bf r}';\varepsilon)
\nonumber\\&
	+
	{\bf G}_0({\bf r},{\bf r_0};\varepsilon)
	\biggl(
		U^{-1}-{\bf G}_0(\varepsilon)
	\biggr)^{-1}
	{\bf G}_0({\bf r_0},{\bf r}';\varepsilon).
\end{align}
It should be highlighted that scattering off the impurity potential $u$ leads to the emergence of a resonance near the Dirac point,
where the position and width of the impurity resonances strongly depend on potential strength, which provides a mechanism for breaking of the electron-hole symmetry. The magnetic scattering potential can be regarded as comprising both repulsive and attractive scattering potentials, one for each spin channel \cite{biswas}. Therefore, a pure magnetic scattering potential preserves electron-hole symmetry, which has a significant influence on the non-linear RKKY interaction, as we shall see below.
After some algebra (see Supplemental material\cite{Supplement}), the non-linear spin susceptibility tensor can be written as
\begin{align}
\label{eq:nonchi}
\boldsymbol{\chi}({\bf{r}},{\bf{ r}}^\prime)=&
-\text{Im}
	\int_{-\infty}^{\varepsilon_F} \frac{d\varepsilon}{2\pi}
			\,  \text{Tr}\:\Big[\frac{
				\boldsymbol{\sigma}{\bf G}_0({\bf r},{\bf r}';\varepsilon)
				\boldsymbol{\sigma}{\bf G}_0({\bf r}',{\bf r};\varepsilon)
			}{1-2gu+g^2u^2-g^2m^2}\Big]
	,
\end{align}
where $g(\varepsilon)={\rm Tr}\;\big[\int d{\textbf{k}}{\bf G}_0(\varepsilon,\textbf{k})\big]/2$.  

As mentioned in Ref. \citenum{mhf}, the RKKY interaction in $3$D TI is strongly direction dependent. By redefining the spin variable according to $\tilde{\bf S}_m=(S_{mx} \sin\varphi_R, S_{my} \cos\varphi_R, S_{mz})$
where $\varphi_R$ is the polar angle of the relative distance between impurities, the effective RKKY Hamiltonian assumes the form
\begin{align}
\label{HRKKY}
H_\text{RKKY}=&
	J_H {\bf S}_1\cdot {\bf S}_2+{\bf J}_\text{DM}\cdot (\tilde{\bf S}_1\times \tilde{\bf S}_2)
\nonumber\\&
	+J_I (\tilde{\bf S}_1\cdot \tilde{\bf S}_2+\tilde{S}_{1x}\tilde{S}_{2y}+\tilde{S}_{1y}\tilde{S}_{2x})
.
\end{align}
for which three kinds of pairing between impurities appear with coefficients: H-type, $J_H$, DM-type, ${\bf J}_\text{DM}=J_\text{DM}(1,-1,0)$ and I-type, $J_I$. See Supplemental Material \cite{Supplement} for details.
\section{Results}
Within the linear response theory, at zero Fermi energy the RKKY interaction for 2D Dirac materials decays as $R^{-3}$ with unchanged sign, in contrast to other 2D material for which it decays as $R^{-2}$.
Moreover, the DM-type interaction is proportional to the spin-orbit coupling and its sign depends on the helicity. Hence, due to the electron-hole symmetry in the TI and opposite helicity in conduction and valance bands, the DM-type coupling is an odd function of the Fermi energy and, hence, vanishes at the Dirac point \cite{nagaosa302}.
In the following, we present the corrections to the RKKY interaction induced by the scattering off the magnetic impurity and the implications thereof. In this paper, the energies are scaled by the band cut-off, $\Lambda=1$ eV, and $m$, $u$ by $\Lambda\lambda^2$. Here, $\lambda$ is the short range cut-off introduced by $ \lambda\equiv\hbar v_F/\Lambda$ which scales the distance ($R$) between impurities. In all figures, three couplings, $J_i$, are presented in units of $ \big(4\,\pi J_c \, \Lambda^{-2}\big)^2 $.\\
\begin{figure*}[t]
\includegraphics[width=0.99\textwidth]{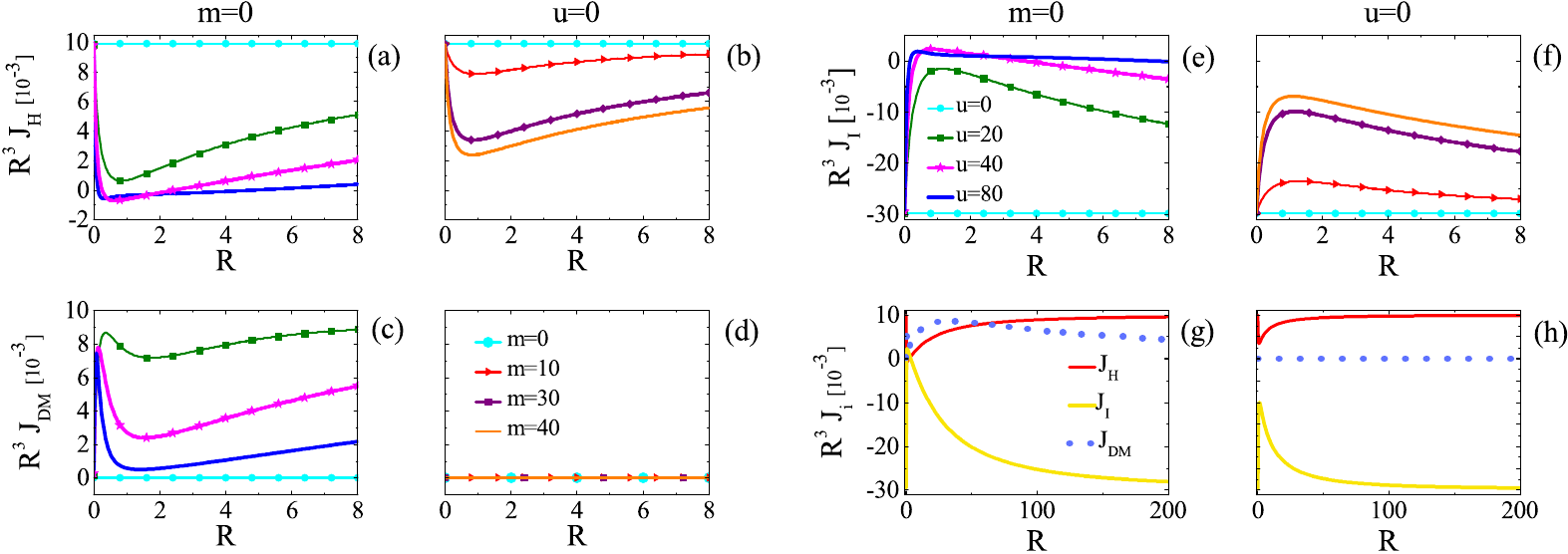}
\caption{\small{(Color online)  Different terms of the RKKY interaction multiplied by $R^3$  as a function of the distance between two impurities at zero doping, $ \varepsilon_F=0 $. In panels (a,c,e) $ m=0 $ and panels (b,d,f), $ u=0 $ in which the magnitudes of $u /{\bf m}$ displayed in the legend of panel (e)/(d) are applied for panels (a, c)/(b, f) respectively while in panel (g), $ m=0 $ and $u=40$ and panel (h), $m=30$ and $ u=0 $. }}
\label{fig.Js_R_Ef0}
\end{figure*}
The spatial dependence of $J_i$, $i=H,I,DM$ is presented in Fig. \ref{fig.Js_R_Ef0} for short (a) -- (f) and long distances (g) -- (h), where we plot the interaction for different values of $u$ (${\bf m}=0$)  and ${\bf m}$ ($u=0$).
The linear response results ($u=0$, ${\bf m}=0$) are included for reference and display a strictly cubic spatial decay as well as vanishing DM contribution.
The impurity scattering, substantially modifies the simply cubic decay of the RKKY interaction as it locally changes the doping of the system. First, we notice that a finite $u$, Fig. \ref{fig.Js_R_Ef0} (a), (c), (e), (g), leads to that all contributions acquire a non-monotonic spatial dependence with strong variation near the impurity. Second, there is a finite range ($2<R\lesssim8$) of nearly quadratic decay for all interactions. Third, by increasing scattering potential $u$, the H and I contributions change sign near the impurity. This behaviour is equivalent to a transition between FM and AFM phases. Fourth, although collinear contributions decrease in amplitude as $u$ is increased, the DM-type interaction becomes the dominating contribution for large $u$, which is expected to have severe implications on the effective magnetic field exerted by the magnetic impurities on the TI surface states.
Fifth, the spatial decay of the impurity resonances leads to that the non-linearity vanishes for large distances, such that the interaction approaches the linear response result (see Fig. \ref{fig.Js_R_Ef0} (g), (h)).
This is consistent with previous \textit{ab-initio} results \cite{PhysRevB.89.165202}, where impurities hosting \textit{d}-electrons were found to reduce the effective exchange interaction.

\begin{figure}[b]
\includegraphics[width=\columnwidth]{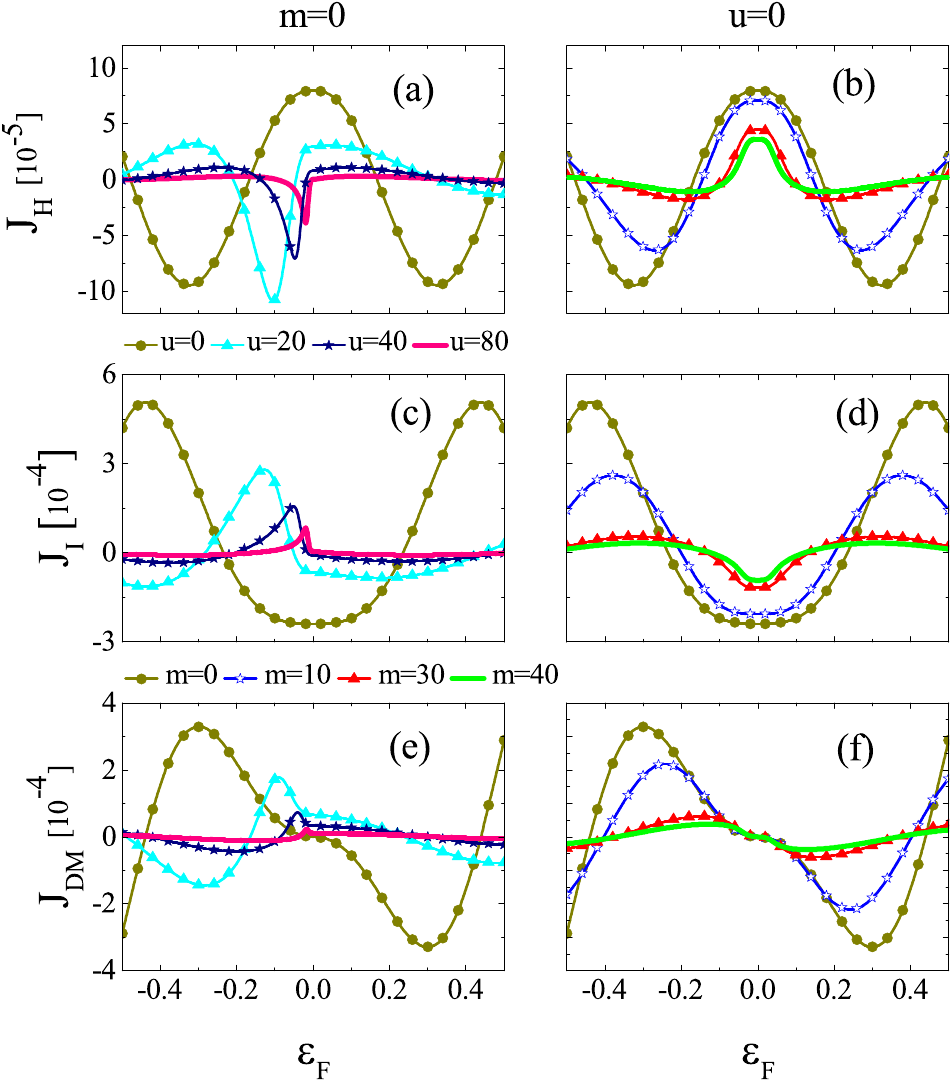}
\caption{\small{(Color online)  Different terms of the RKKY interaction as a function of the Fermi energy for distance $ R=2 $ between impurities and different values of impurity potential $u, m $. The used values of $u /{\bf m}$ for panels (a, c, e)/(b, d, f) are found below panel (a)/(b).}}
\label{fig.JsEf_R5}
\end{figure}

It should be noticed, while the combination of a finite magnetic and vanishing non-magnetic scattering potential, see Figs.\ref{fig.Js_R_Ef0} (b), (d), (f), (h), yields a vanishing DM-type contribution, the non-monotonic spatial dependence of $J_H$ and $J_I$ remain as before. In this limit, one can expect an FM formation of the magnetic impurities.
Although some of these behaviours are established also for ${\bf m}=0$, the effect of the magnetic potential is smaller than the $u$ term. In particular, the sign of the interaction remains intact with growing $|{\bf m}|$.
  
While the linear response theory yields a vanishing DM-type contribution, Fig. \ref{fig.Js_R_Ef0} (c) shows that it is non-negligible whenever the non-magnetic scattering potential is finite. Note that a mere magnetic scattering potential ($u=0$, ${\bf m}\neq0$) is not sufficient to provide a finite DM-type interaction (Fig. \ref{fig.Js_R_Ef0} (d)). We attribute this property to that a purely magnetic scattering potential preserves the electron-hole symmetry present in Dirac materials. The non-magnetic scattering potential breaks this symmetry by introducing local doping which leads to a finite $J_\text{DM}$. We expect that this property can be used in spintronics devices with electrical tunability. The plots in Fig. \ref{fig.Js_R_Ef0} suggest that the scattering potential $u$ can make this contribution dominating over $J_H$ and $J_I$, something which may have an impact on the functionality.

At finite doping, $\varepsilon_F \neq 0$, the interaction parameters acquire an oscillating dependence on the Fermi wave vector and distance $R$ between the spin moments. The plots in Fig. \ref{fig.JsEf_R5} show the dependencies of $\varepsilon_F$ for varying strengths of the scattering potentials, where the linear response ($\sin2k_FR$) result is included for reference (dark yellow curve). The plots in the left panels clearly show the electron-hole symmetry breaking caused by a finite $u$ while the right panels show that it is preserved under purely magnetic scattering potentials.
Importantly, the scattering potential changes the oscillations and the sign of all terms in a wide range of energies, suggesting that non-linearity terms cannot be neglected without loosing accuracy in the theoretical description.
It should be noticed that both $u$ and ${\bf m}$ tends to reduce the magnitude of the RKKY interactions. However, for a wide range $|\varepsilon_F|<100$ meV, that non-magnetic impurity scattering enhances the DM-type contribution while $J_H$ and $J_I$ are suppressed, consistent with the effect of the impurity scattering of the spatial decays.
\begin{figure}[t]
\includegraphics[width=1\columnwidth]{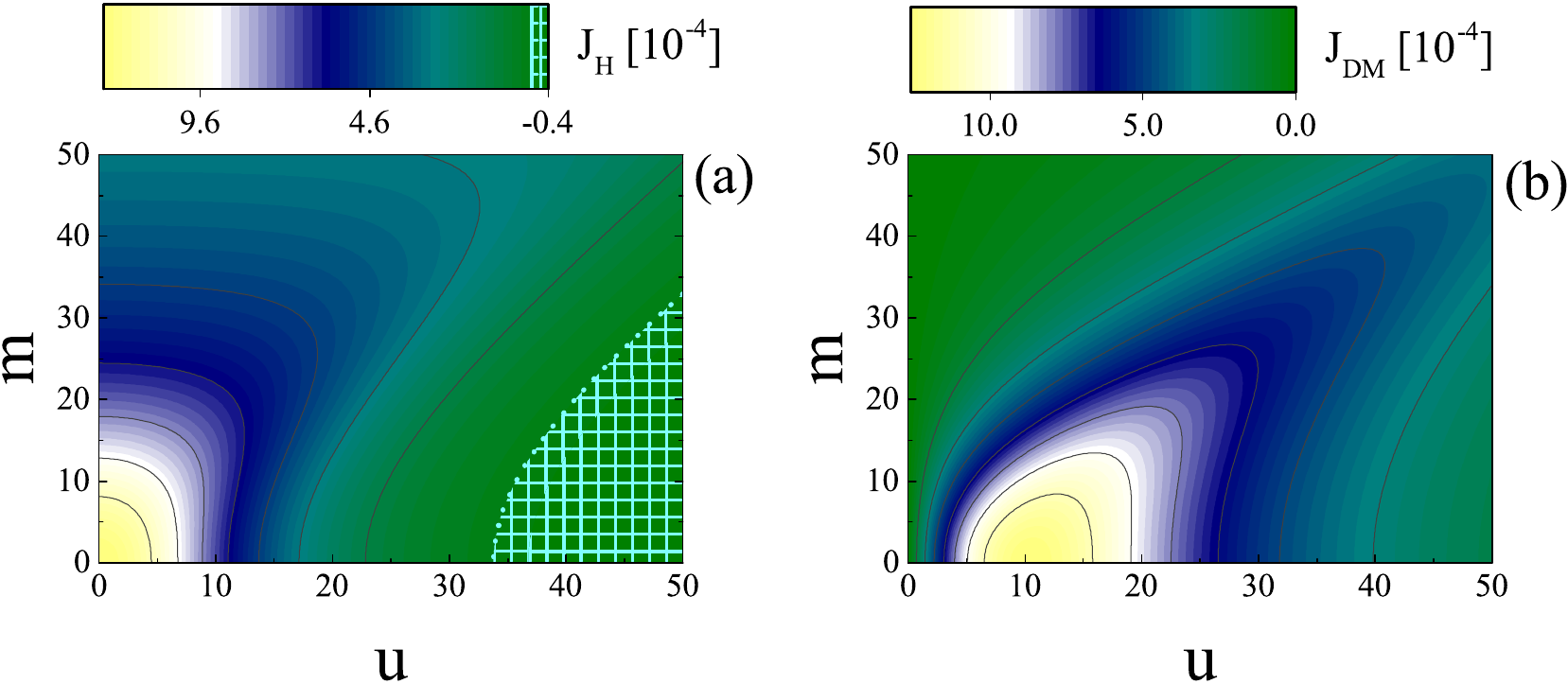}
\caption{\small{(Color online) The contour plot of the RKKY couplings (a) $J_H$ and (b) $J_{DM}$  in the plane of impurity potential terms $u,m $ at zero doping, $ \varepsilon_F=0 $ and $ R=2 $. }}
\label{fig.JHDMum}
\end{figure}
The simultaneous effect of $ u $ and $ {\bf m} $ terms on the isotropic and anti-symmetric anisotropic components of the RKKY interaction is plotted in Fig. \ref{fig.JHDMum}, which shows the parameters (a) $J_{H}$ and (b) $J_{DM}$ at $R=2$, and at zero doping ($\varepsilon_F=0$). The rastered region in panel (a), indicates a sign change of the H type spin-spin interaction such that anti-parallel alignment of the spins is favoured over a parallel one, suggesting the possibility of a magnetic phase transition. This region is shifted to higher values of $u$ with increasing distance $R$, reflecting the fact that the impurity states decay with the distance such that impurity coupling approaches the linear behaviour. The DM-type contribution vanishes at $u=0$ line for all values of $|{\bf m}|$ (Fig. \ref{fig.JHDMum} (b)). Moreover, the parameter $J_\text{DM}$ is a non-monotonic function of $u$, with a wide peak around a finite value of $u$. At zero Fermi energy, this parameter is generated by the electron-hole symmetry breaking introduced by a finite $u$. However, with increasing $u$,  the impurity resonance approaches the Dirac point while its life-time increases which leads to a reduced electron--hole asymmetry. The competition between different types of coupling opens possibilities to optimize the properties of the magnetic interactions, the resulting effective magnetic field, and the influence of the magnetic impurities on the electronic structure of surface state of the host TI. It has previously been shown that the ratio between the magnetic and non-magnetic scattering potentials strongly influences the possibilities for a gap opening near the Dirac point \cite{jonas-filling}. The results obtained here, moreover, suggest that a small ratio $|{\bf m}|/u$ would favour non-collinear configurations of the magnetic impurities as the interaction between these are dominated by $J_\text{DM}$. It is then expected that the $z$-component of the total magnetic field generated by the magnetic impurities is strongly reduced, such that the size of the density gap around the Dirac point is significantly diminished.

\begin{figure}[t]
\includegraphics[width=1\columnwidth]{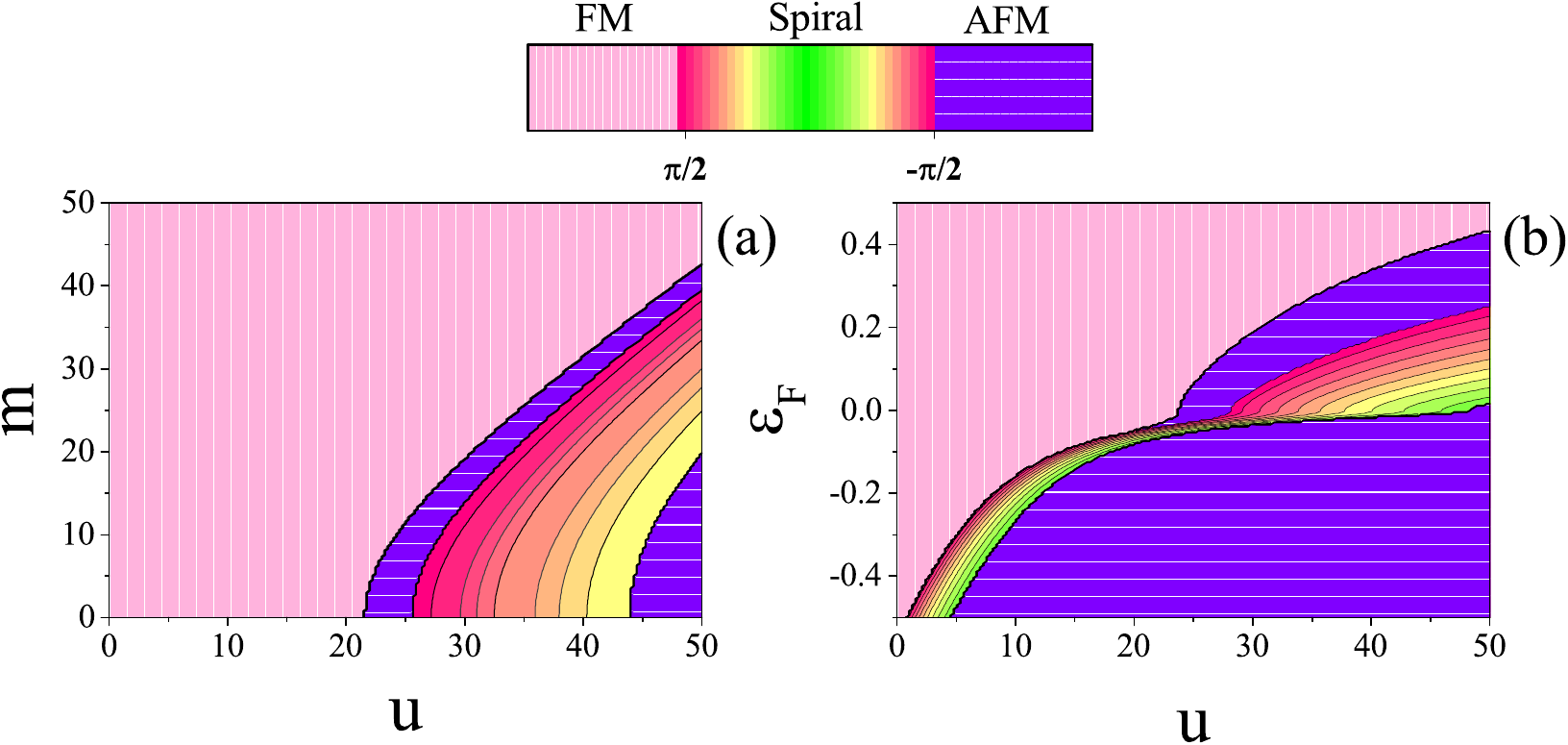}
\caption{\small{(Color online) The phase diagram of the situation of two magnetic impurities with respect to each other for  (a) $\varepsilon_F=0$ with respect to $u,m$ and (b) for $\textbf{m}=0$ in the plane of $u,\varepsilon_F$. In both cases, we assumed the relative position of the impurities to be $\textbf{R}=(2,0)$. }}
\label{figphase}
\end{figure}

In order to better understanding of the impact of the non-linear interaction and the competition between the different types of coupling, we investigate the ordering of two impurities in Fig. \ref{figphase}. Following references \cite{fariborz-mos2, fariborz-silicene1}, we find that in presence of the DM-type term, the magnetic moments become non-collinear, hence, defining an angle $\phi=\arctan(J_{DM}/J_{H})$ (here referred to as spiral order) between each other in the plane perpendicular to $J_\text{DM}$. However, this phase does not necessarily correspond to the ground state of the system and the spins can also be FM or AFM aligned.
Figure \ref{figphase} (a), shows the phases of two impurities located at ${\bf r}_1=(0,0)$ and ${\bf r}_2=(2,0)$, at $\varepsilon_F=0$ (at which value the linear response theory predicts an FM ground state). The figure shows that inclusion of impurity scattering opens a wide range of the $(u,|{\bf m}|)$-plane in which the impurities are either in AFM or in non-collinear configuration, where in the latter case $-\pi/2<\phi<\pi/2$. Figure \ref{figphase} (b) illustrates the ordering of the two impurities as a function of $u$ and $\varepsilon_F$, for $|{\bf m}|=0$. The asymmetric behaviour about $\varepsilon_F=0$ is expected due to broken electron--hole symmetry caused by the potential scattering. The details of derivation of the phases, can be found in the supplementary material.
\section{Conclusion}
In conclusion, we have revised the theory for indirect spin-spin interactions in Dirac materials by including the influences of the impurity scattering on the local electronic structure. In particular, we have studied the effect of impurity states on the RKKY interaction mediated by the surface states of $3$D TIs and found that the impurity states substantially affect the RKKY interaction and intensively modify the picture obtained from linear response theory.
In particular, the emergence of impurity resonances from both magnetic and non-magnetic scattering potentials tend to reduce the H and I contributions and may even lead to sign changes of the interactions. In contrast, the DM-type interaction at zero doping, predicted to vanish in linear response theory, it is not only finite but becomes the dominating interaction for large ratios between the non-magnetic and magnetic scattering potentials. Our results are shown to be stable under finite doping. Based on our results, we predict that the deepened insight to the magnetic interactions may revise the picture concerning the possibilities to create density gaps around the Dirac point using magnetic impurities. For the sake of clarity, our results are presented for a wide range of parameters, which largely overlap with accessible experimental values. More precisely, $\lambda\simeq2.7$ \AA\ for Bi$_2$Se$_3$ which suggests that the distances we have chosen are relevant to distances between impurities in magnetic TIs. The range of Fermi energy has been also tested in practice \cite{KOU201534}. Following Ref. \citenum{PhysRevB.96.235444}, which compared the LDOS with scanning tunnelling spectroscopy results, one can find that the strength of impurity potentials, $m$ and $u$ can vary between $0.3-3$ and $3-30$, respectively. This shows the practical importance of $u$ for different magnetic impurities.

\section{Acknowledgment}
F.P. thanks the fruitful discussion with A.V. Balatsky and A. Qaiumzadeh. M.Sh. thanks Saeed Amiri for his technical support. H.C. thanks the International Center for Theoretical Physics (ICTP) for their hospitality and support during the regular associate program of the center. J.F. acknowledges support from Vetenskapsr\aa det.
\bibliography{RefGRKKY}

\end{document}